\documentstyle[epsfig,mprocl,12pt]{article}

\begin{document}
\bibliographystyle{unsrt}

\vbox {\vspace{6mm}}

\begin{center}
{\large \bf $\Lambda\Lambda$-$\Xi$N 
coupling effects in light hypernuclei}\\[7mm]
Khin Swe Myint, Shoji Shinmura and Yoshinori Akaishi\\ [3mm]
{\it Department of Physics, Mandalay University, Mandalay, 
Union of Myanmar}\\ 
{\it Department of Information Science, Gifu University, Gifu 501-1193, 
Japan}\\
{\it Institute of Particle and Nuclear Studies, KEK, Tsukuba 305-0801, 
Japan}\\ [5mm]
\end{center}

\begin{abstract}
The significance of $\Lambda\Lambda$-$\Xi$N coupling in 
double-$\Lambda$ hypernuclei has been studied. The Pauli suppression effect 
due to this coupling in $^6_{\Lambda\Lambda}$He has been found to be 
0.43 MeV for the coupling strength of the NSC97e potential. 
This indicates that the free-space $\Lambda\Lambda$ 
interaction is stronger by about $5^{\circ}$ phase shift  
than that deduced from the empirical data of 
$^6_{\Lambda\Lambda}$He without including the Pauli suppression 
effect. In $^5_{\Lambda\Lambda}$He and $^5_{\Lambda\Lambda}$H, an 
attractive term arising from $\Lambda\Lambda$-$\Xi$N conversion is 
enhanced by the formation of an alpha particle in intermediate $\Xi$ 
states. According to this enhancement, we have found that the 
$\Lambda\Lambda$ binding 
energy ($\Delta B_{\Lambda\Lambda}$) of $^5_{\Lambda\Lambda}$He is about 
0.27 MeV larger than that of $^6_{\Lambda\Lambda}$He for the NSC97e 
coupling strength. This finding deviates from a general picture that 
the heavier is the core nucleus, the larger is $\Delta 
B_{\Lambda\Lambda}$.  
\end{abstract}

\section{Introduction}

The significant role of $\Lambda$N-$\Sigma$N coupling in $s$-shell  
single-$\Lambda$ hypernuclei and 
in neutron matter with a $\Lambda$ has been presented by us in previous 
papers. \cite{1,2} It is also our interest to find out the effect of 
$\Lambda\Lambda$-$\Xi$N coupling in double-$\Lambda$ hypernuclei, 
since the mass difference between the $\Lambda$ and $\Xi$ channels is 
only 28 MeV, which is smaller than that of the $\Lambda$ and $\Sigma$ 
channels. Recently, in the KEK E373 emulsion counter hybrid experiment a 
$^6_{\Lambda\Lambda}$He, known as ``Nagara'' event, \cite{3} was observed 
unambiguously.   
Filikhin and Gal \cite{4} have carried out Faddeev-Yakubovsky  
three- and four-body calculations 
to analyze $^6_{\Lambda\Lambda}$He 
and other double-$\Lambda$ hypernuclei with various Nijmegen OBE YN 
potential models. However, we notice that the Pauli suppression effect 
 has not 
been included in their calculations,  
which motivates us to investigate this effect on the  
$^6_{\Lambda\Lambda}$He, $^5_{\Lambda\Lambda}$He and 
$^5_{\Lambda\Lambda}$H binding energies and on the deduction of 
$\Lambda\Lambda$ interaction in free space. In this paper we reveal
 an appreciable Pauli suppression effect of  
$\Lambda\Lambda$-$\Xi$N coupling in $^6_{\Lambda\Lambda}$He and a 
significant enhancement effect in $^5_{\Lambda\Lambda}$He.

\section{$\Lambda$-nucleus and $\Lambda\Lambda$ potentials}

We treat the double-$\Lambda$ hypernuclei as 
$\Lambda$+$\Lambda$+core nucleus three-body systems, and first 
prepare the necessary $\Lambda$-nucleus and $\Lambda\Lambda$ 
potentials. 
A hyperon-nucleon potential, D2, 
which essentially solves the overbinding problem in $s$-shell $\Lambda$ 
hypernuclei, \cite{1} is used to obtain  
$\Lambda$-nucleus folding potentials by the Brueckner-Hartree-Fock method. 
They are slightly 
modified \cite{5} so as to reproduce the experimental binding energies of 
the respective hypernuclei,
$^4_{\Lambda}$H, $^4_{\Lambda}$He and $^5_{\Lambda}$He, and 
are expressed in the following two-range Gaussian form:
\begin{equation}
V_{\Lambda -\rm {nucleus}}(r) = \sum_{i=1}^2 V_i e^{-(\frac r{\mu_i})^2} 
\end{equation}
with the parameters given in table 1.
 
\begin{table}[htb]
\begin{center}
\caption{\label{tab:2gaussian}
 Strength and range parameters of $s$-shell $\Lambda$-nucleus 
potentials.} 
\begin{tabular}{ccccccc} \\
\hline \hline  \\
  & $V_1$ & $V_2$ & $\mu_1$ & $\mu_2$ & $B_{\Lambda}$ & 
  $B_{\Lambda}^{\rm exp}$ \\
  &[MeV]&[MeV]&[fm]&[fm]&[MeV]&[MeV]
   \\ 
\hline \\
$V_{\Lambda-{\rm h}}(0^+)$ & 58.1 & -78.4 & 1.40  & 1.72 & 2.39 & 
2.39$\pm$ 0.03  \\ \\
$V_{\Lambda-{\rm h}}(1^+)$ & 71.0 & -81.0 & 1.40  & 1.72 & 1.23 & 
1.24$\pm$ 0.04  \\ \\
$V_{\Lambda-{\rm t}}(0^+)$ & 58.1 & -76.8 & 1.40 & 1.72 & 2.05 & 
2.04$\pm$ 0.04  \\ \\
$V_{\Lambda-{\rm t}}(1^+)$ & 71.0 & -79.6 & 1.40 & 1.72 & 1.01 & 
1.00$\pm$ 0.04  \\ \\ 
$V_{\Lambda-\alpha}$($l$=0)  &
 91.0 & -95.0 & 1.30 & 1.70 & 3.12 & 3.12$\pm$ 0.02   \\ \\
\hline \hline  
\end{tabular}
\end{center}
\end{table}

A single-channel $\Lambda\Lambda$ interaction in free space is derived    
from the diagonal and off-diagonal terms of a strangeness $S$ = -2 coupled 
interaction as follows:
\begin{equation}
V_{\Lambda\Lambda}^{\rm sc} = 
V_{\Lambda\Lambda,\Lambda\Lambda} - 
V_{\Lambda\Lambda,\Xi^- {\rm p}}\frac 1{\Delta E}V_{\Xi^- {\rm 
p},
\Lambda\Lambda} - 
V_{\Lambda\Lambda,\Xi^0 {\rm n}}\frac 1{\Delta E}V_{\Xi^0 {\rm n},
\Lambda\Lambda}.
\end{equation}
 The diagonal term $V_{\Lambda\Lambda,\Lambda\Lambda}$ and the 
 coupling terms $V_{\Lambda\Lambda,\Xi {\rm N}}$ are 
 Shinmura's potentials which are phase-shift equivalents to the Nijmegen 
 soft-core NSC97e \cite{6} and the Nijmegen hard-core NHC-D \cite{7} and 
 NHC-F \cite{8} potentials; these are in Gaussian 
 forms with the parameters given in table 2. A hard-core radius 
 parameter $r_{\rm c}$ = 0.56 (0.53) fm, which is common to all 
 NN and YN $^1S_0$ states, is used in the NHC-D (F) potential. 
 $\Delta E$ of eq. (2) is the operator in intermediate propagation, 
 and the single-channel 
 $\Lambda\Lambda$ interaction becomes a 
 non-local potential. For convenience of 
 practical use, we substitute for it a phase-shift equivalent 
 local potential in a closure approximation in which the $\Delta E$ is 
 replaced by an averaged value $\bar {\Delta E}$.
 The resultant local 
 single-channel interaction is given by 
 \begin{equation}
 V_{\Lambda\Lambda}^{\rm sc}(r) = \sum_{i=1}^3 V_i e^{-(\frac r{\mu_i})^2}
 \end{equation}
 with $\mu_{1}$ = 0.35 fm, $\mu_2$ = 0.85 fm and $\mu_3$ = 0.60 fm.
 
\begin{table}[htb]
\begin{center}
\caption{\label{tab:Shinmura}
Strength and range parameters of Shinmura's potentials, where 
$V_{\Lambda\Lambda,\Xi \rm N} \equiv 
V_{\Lambda\Lambda,\Xi^- \rm p} = V_{\Lambda\Lambda,\Xi^0 \rm n}$ is 
given in charge base, not in isospin base.} 
\begin{tabular}{ccccc} \\
\hline \hline  \\
  &$V_1$ & $V_2$ & $\mu_1$ & $\mu_2$ \\
  &[MeV]& [MeV]& [fm]& [fm]
   \\ 
\hline \\
$V_{\Lambda\Lambda,\Lambda\Lambda}$(D) & 22912 & -384.7 & 0.35  & 
0.85  \\ \\
$V_{\Lambda\Lambda,\Xi \rm N}$(D) & &50.9 &  & 0.85   \\ \\
\hline \\
$V_{\Lambda\Lambda,\Lambda\Lambda}$(97e) & 18927 & -286.8 & 0.35  & 
0.85  \\ \\
$V_{\Lambda\Lambda,\Xi \rm N}$(97e) & &108.6 &  & 0.85   \\ \\
\hline \\
$V_{\Lambda\Lambda,\Lambda\Lambda}$(F) & 14080 & -198.6 & 0.35  & 
0.85  \\ \\
$V_{\Lambda\Lambda,\Xi \rm N}$(F) & &143.7 &  & 0.85   \\ \\
\hline \hline  
\end{tabular}
\end{center}
\end{table}

First, we construct the local single-channel $V_{\Lambda\Lambda}^{\rm sc}$ 
by adjusting  $\bar {\Delta E}$$>$0 to reproduce the scattering 
parameters of the NSC97e $S$ = -2 interaction. We found that
 $\bar {\Delta E}$ = 137.6 MeV reproduces the scattering length, 
 $a_{\Lambda\Lambda}$ = -0.50 fm, and the effective range,
 $r_{\Lambda\Lambda}$ = 8.41 fm.  
  This local single-channel 
 $\Lambda\Lambda$ potential is given in table 3 as 
 $V_{\Lambda\Lambda}^{\rm e}$ and it well reproduces the phase shifts 
 of the non-local potential of eq. (2) in the region of $E_{\rm c.m.}$ = 
 0 $\sim$ 15 MeV. 
 A measure of $\Lambda\Lambda$ interaction is given by 
 \begin{equation}
 \Delta B_{\Lambda\Lambda}(^{\rm A}_{\Lambda\Lambda}{\rm X}) = 
 B_{\Lambda\Lambda}(^{\rm A}_{\Lambda\Lambda}{\rm X}) - 2 
 B_{\Lambda} (^{\rm A-1}_{\Lambda}{\rm X}).
 \end{equation}
 The $\Lambda\Lambda$ binding energy of 
 $^6_{\Lambda\Lambda}$He is found to be $\Delta B_{\Lambda\Lambda}$
 = 0.64 MeV which is about 0.4 MeV smaller than that of the 
Nagara event data, \cite{3} 
$\Delta B_{\Lambda\Lambda}$ = $1.01\pm 0.20^{+0.18}_{-0.11}$ MeV,
when we employ this single-channel $V_{\Lambda\Lambda}^{\rm e}$ 
and $V_{\Lambda-\alpha}$ 
 of table 1. Because this result is in good agreement with that 
 of Filikhin and Gal, \cite{4} 
 one may think that we have reached the same conclusion that NSC97e is 
 an appropriate model for reproducing the recent experimental 
 value of $\Delta B_{\Lambda\Lambda}(^6_{\Lambda\Lambda}$He). However, 
 since the effect of the $\Lambda\Lambda$-$\Xi$N coupling is already  
included implicitly in the single-channel 
$V_{\Lambda\Lambda}^{\rm e}$, we must consider the  Pauli 
suppression effect in $^6_{\Lambda\Lambda}$He, 
where all of the 0$s$ states are forbidden to a nucleon 
converted from the $\Lambda\Lambda$-$\Xi$N coupling. 
Since the Pauli suppression 
effect would cause a serious change in our results as will be discussed 
later, the above conclusion about NSC97e should be altered.
Before introducing the Pauli suppression, we modify the 
$V_{\Lambda\Lambda}^{\rm e}$ by adjusting the long-range part of 
$V_{\Lambda\Lambda,\Lambda\Lambda}$ in eq. (2) to fit the value 
$\Delta B_{\Lambda\Lambda}$ = 1.01 MeV recommended from the Nagara event.
 We call this fitted $\Lambda\Lambda$ interaction
 $V_{\Lambda\Lambda}^{\rm e1}$. 
 It is noticed that this is not $\Lambda\Lambda$ 
 interaction in free space. 
 Carr, Afnan and Gibson \cite{9} discussed the 
significance of the Pauli suppression effect in deducing the 
$\Lambda\Lambda$ interaction in free space 
from the experimental $\Lambda\Lambda$ 
binding energies of double-$\Lambda$ hypernuclei.

 The Pauli suppression effect in $^6_{\Lambda\Lambda}$He is given by
\begin{equation}
\Delta V_{\rm Pauli} = V_{\Lambda\Lambda,\Xi^- {\rm p}}\frac {P_{\alpha}}
{\Delta M} V_{\Xi^- 
{\rm p},\Lambda\Lambda} + 
V_{\Lambda\Lambda,\Xi^0 {\rm n}}\frac {P_{\alpha}}{\Delta M} V_{\Xi^0 
{\rm n},\Lambda\Lambda},
\end{equation}
where $P_{\alpha}$ is the projection operator on the 0$s$ nucleon 
states in an $\alpha$ particle. 
 Here, we also restrict the $\Xi$ states to 0$s$ to estimate 
 the minimum effect of the Pauli suppression. Then,  
 \begin{equation}
 \Delta M = \frac {M_{\Xi^0} + M_{\Xi^-}}2 
 + \frac {M_{\rm p} + M_{\rm n}} 2 - 2 M_{\Lambda}
 + 2 B_{\Lambda} (^5_{\Lambda}{\rm He}) = 32.0~{\rm MeV}, 
 \end{equation}
 where we neglect the $\Lambda\Lambda$ binding energy 
 $\Delta B_{\Lambda\Lambda}$ since it is nearly cancelled by binding 
 energy of the $\Xi$. \cite{10} 
 $\Delta V_{\rm Pauli}$ is expressed as a two-body non-local potential with 
\begin{equation}
P_{\alpha} = \mid 0s 0s\rangle _{\alpha}¥_{\alpha}\langle 0s 0s \mid
\end{equation} 
which is given in appendix A.
We then define the $V_{\Lambda\Lambda}^{\rm e2}$ as 
\begin{equation}
V_{\Lambda\Lambda}^{\rm e2} = V_{\Lambda\Lambda}^{\rm e1} 
- \Delta V^{\rm e}_{\rm Pauli},
\end{equation} 
where the superscript in the second term means 
that the NSC97e coupling strength is used in eq. (4).
In other words, 
 $V_{\Lambda\Lambda}^{\rm e2}$ + 
$\Delta V^{\rm e}_{\rm Pauli}$ produces $\Delta 
B_{\Lambda\Lambda}(^6_{\Lambda\Lambda}{\rm He})$ = 1.01 MeV, whereas 
 $V_{\Lambda\Lambda}^{\rm e2}$ alone gives 1.44 MeV as shown later. This 
 $V_{\Lambda\Lambda}^{\rm e2}$ is the single-channel $\Lambda\Lambda$ 
 interaction in free space, which reproduces the Nagara event data,
  and is more attractive than the fitted 
  $V_{\Lambda\Lambda}^{\rm e1}$ 
  and the original $V_{\Lambda\Lambda}^{\rm e}$.
 Similarly, the single-channel $\Lambda\Lambda$ 
 interactions $V_{\Lambda\Lambda}^{\rm D}$,    
 $V_{\Lambda\Lambda}^{\rm D1}$, $V_{\Lambda\Lambda}^{\rm D2}$, 
 $V_{\Lambda\Lambda}^{\rm F}$, 
  $V_{\Lambda\Lambda}^{\rm F1}$ and  $V_{\Lambda\Lambda}^{\rm F2}$ 
  are constructed from the models NHC-D and NHC-F, respectively.
  The strength parameters of the various $\Lambda\Lambda$ 
 interactions are given in table 3.

\begin{table}[htb]
\begin{center}
\caption{\label{tab:Lambda-Lambda}
Strength parameters of the $\Lambda \Lambda$ potentials in 
units of MeV.} 
\begin{tabular}{cccc} \\
\hline \hline  \\
  &$V_1$ & $V_2$ & $V_{3}$     \\ \\
\hline \\
$V_{\Lambda\Lambda}^{\rm D}$ & 22912 & -384.7  & -73.0    \\ \\
$V_{\Lambda\Lambda}^{\rm D1}$ & 22912 & -356.5  & -73.0   \\ \\
$V_{\Lambda\Lambda}^{\rm D2}$ & 22912 & -361.9  & -73.0   \\ \\
$V_{\Lambda\Lambda}^{\rm D3}$ & 22912 & -418.7  & -73.0   \\ \\
\hline   \\
$V_{\Lambda\Lambda}^{\rm e}$ & 18927 & -286.8 & -171.4  \\ \\
$V_{\Lambda\Lambda}^{\rm e1}$ &18927 & -311.2  & -171.4  \\ \\
$V_{\Lambda\Lambda}^{\rm e2}$ &18927 & -336.4  & -171.4  \\ \\
$V_{\Lambda\Lambda}^{\rm e3}$ & 18927 & -391.3  & -171.4   \\ \\
\hline   \\
$V_{\Lambda\Lambda}^{\rm F}$ & 14080 & -198.6 & -296.2  \\ \\
$V_{\Lambda\Lambda}^{\rm F1}$ &14080 & -246.3  & -296.2  \\ \\
$V_{\Lambda\Lambda}^{\rm F2}$ &14080 & -291.4  & -296.2 \\ \\
$V_{\Lambda\Lambda}^{\rm F3}$ & 14080 & -343.2  & -296.2   \\ \\
\hline \hline 
\end{tabular}
\end{center}
\end{table} 

\section{Results and discussions}

We perform the three-body calculations on $\Lambda$ + $\Lambda$ + 
$\alpha $,
$\Lambda$ + $\Lambda$ + h and $\Lambda$ + $\Lambda$ + t systems, where 
the total wave function of the systems is expanded in terms of the 
Gaussian wave functions, which are spanned over the three 
rearrangement channels in Jacobi coordinates. \cite{11}  
Odd state $\Lambda$-$\alpha$ interaction, which is not well determined, 
is assumed to be zero because its contribution is rather small. 
A detailed account of our three-body calculation method and 
its accuracy are discussed in appendix A.
 The calculated values of the $\Lambda\Lambda$ binding energies with 
 various $\Lambda\Lambda$ potentials are summarized in table 4. 

\subsection{$\Lambda\Lambda$-$\Xi$N coupling effects in 
$^6_{\Lambda\Lambda}$He}¥
  
 We have derived the single-channel 
 $\Lambda\Lambda$ interactions based on Shinmura's $S$ = -2 
 interactions, which are phase-shift equivalents to the 
 NSC97e, NHC-D and NHC-F models.
 The free-space $\Lambda\Lambda$ interactions, $V_{\Lambda\Lambda}^{\rm 
 D2}$, 
 $V_{\Lambda\Lambda}^{\rm e2}$ and $V_{\Lambda\Lambda}^{\rm F2}$,
  give $\Delta 
 B_{\Lambda\Lambda}(^6_{\Lambda\Lambda}{\rm He})$ = 1.10 MeV, 1.44 MeV and 1.89 
 MeV, respectively,   
 and when the Pauli suppression effect is included they reduce to the 
 empirical value, 1.01 MeV. 
Thus, the Pauli suppression effect due to the 
 $\Lambda\Lambda$-$\Xi$N coupling is 0.09 MeV for 
 NHC-D, 0.43 MeV for NSC97e and 0.88 MeV 
for NHC-F in $^6_{\Lambda\Lambda}$He.
The $\Lambda$-$\Lambda$ scattering phase shifts derived from the potentials 
 $V_{\Lambda\Lambda}^{\rm e1}$, 
 $V_{\Lambda\Lambda}^{\rm e2}$ and 
 $V_{\Lambda\Lambda}^{\rm F2}$ are shown in fig. 1. 
The figure indicates that a $12^{\circ}$ phase shift at the maximum is 
obtained for $V_{\Lambda\Lambda}^{\rm e1}$ if 
the Nagara event data is fitted by ignoring the Pauli suppression effect. 
For $V_{\Lambda\Lambda}^{\rm e2}$ and $V_{\Lambda\Lambda}^{\rm F2}$,   
which are fitted by including the Pauli suppression effect, 
the $\Lambda$-$\Lambda$ phase shifts attain $17^{\circ}$ and  
$22^{\circ}$, respectively. 
 The $\Lambda$-$\Lambda$ 
  scattering lengths are $a_{\Lambda\Lambda}^{\rm e1}$ = -0.73 fm,  
   $a_{\Lambda\Lambda}^{\rm e2}$ = -1.03 fm,
    $a_{\Lambda\Lambda}^{\rm F1}$ = -0.67 fm, and 
    $a_{\Lambda\Lambda}^{\rm F2}$ = -1.32 fm, respectively. 
 It should be stressed that when we derive the 
   free-space single-channel $\Lambda\Lambda$ interaction from
    the Nagara-event data, the Pauli effect due 
    to the $\Lambda\Lambda$-$\Xi$N coupling should be consistently taken 
    into account.    
In the case of the NSC97e 
coupling strength, the potential in free space 
$V_{\Lambda\Lambda}^{\rm e2}$ is much more attractive than 
$V_{\Lambda\Lambda}^{\rm e1}$ and $V_{\Lambda\Lambda}^{\rm e}$, the 
latter of which is equivalent to the original NSC97e $S$ = -2 interaction. 
The situation is more pronounced (restrained)
with the NHC-F (NHC-D) interaction, which has a 
stronger (weaker) $\Lambda\Lambda$-$\Xi$N coupling strength mainly 
due to the value 3.33 (0.94) of the $f_{\Xi\Lambda K^*}$ baryon-meson coupling 
constant. 
\cite{12}  

\vspace{-0.0cm}
\begin{figure}[htb]
\centerline{\epsfig{file=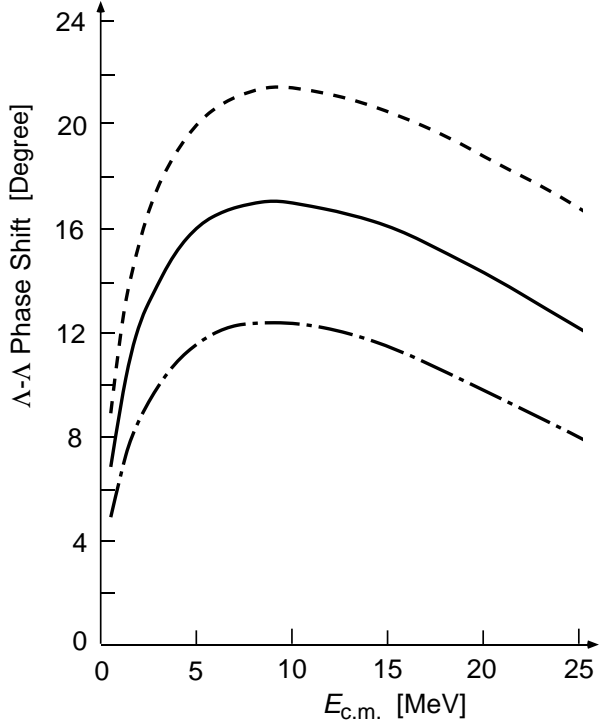,width=12.5cm}}
\vspace{-6.5cm}
\caption{$\Lambda$-$\Lambda$ phase shifts for different cases of the   
$\Lambda\Lambda$-$\Xi$N coupling strength. 
The dot-dashed line is for $V_{\Lambda\Lambda}^{\rm e1}$, the 
solid line is for $V_{\Lambda\Lambda}^{\rm e2}$ and the dashed line is 
for $V_{\Lambda\Lambda}^{\rm F2}$.}
\label{figXi1}
\end{figure}
 
 Filikhin and Gal \cite{4} showed that the 
$\Delta B_{\Lambda\Lambda}$ of $^6_{\Lambda\Lambda}$He with the NSC97e
 $\Lambda\Lambda$ interaction is about 0.5 MeV smaller than that of 
 the Nagara event without the Pauli suppression effect. 
 They discussed that the Pauli 
 suppression effect would be completely cancelled out by a 
 higher partial-wave effect.  
 We have checked 
 the contributions of higher partial waves to the binding energy of 
 $^6_{\Lambda\Lambda}$He. The contribution due to 
 $\Lambda\Lambda$ interaction in $l$ = 1  and higher states  
 is only 0.0005 MeV which is negligibly small. For 
 $\Lambda$-$\alpha$ interaction, the corresponding contribution is 
 0.14 MeV. Our $\Lambda$-$\alpha$ interaction for $l$ = 1, 
 derived from NSC97e, is less attractive than the one for $l$ = 0
 given in table 1, and is parametrized with $V_1$ = 33.4 MeV and 
 $V_2$ = -39.4 MeV on the same ranges. Thus, if we treat the NSC97e case
  in a consistent way, the Pauli suppression 
 effect of 0.43 MeV cannot be cancelled out by the higher partial-wave 
 effect and a net effect of 0.30 MeV remains. 
  
Recently, Nemura {\it et al.} \cite{13}
 pointed out that a large rearrangement effect, due to the presence 
 of $\Lambda$ particle, on the internal energy of $\alpha$ 
  core takes place in $^5_{\Lambda}$He. 
 If we take into account the rearrangement effect, 
 we must reproduce with the $\Lambda\Lambda$ interaction the 
  $\Delta B_{\Lambda\Lambda}$ of the Nagara-event data plus 
  $\Delta B_{\rm rearr}$ of the rearrangement which is approximately 
  estimated to be 1 MeV by Kohno. \cite{14}
   The $\Lambda\Lambda$ 
  interactions fitted to thus increased $\Delta 
  B_{\Lambda\Lambda}+\Delta B_{\rm rearr}(^6_{\Lambda\Lambda}$He) 
  are denoted as 
  $V_{\Lambda\Lambda}^{\rm D3}$, $V_{\Lambda\Lambda}^{\rm e3}$ and 
  $V_{\Lambda\Lambda}^{\rm F3}$, and the respective Pauli suppression 
  effects are 0.12, 0.58 and 1.19 MeV. The maximum value of 
  $\Lambda\Lambda$ phase shift attains $30^{\circ}$ for NSC97e. More 
  careful
  investigation on this rearrangement effect would be done elsewhere.

 \subsection{Characteristic features of $^5_{\Lambda\Lambda}{\rm He}$ 
and $^5_{\Lambda\Lambda}{\rm H}$ } 

We then investigate the change of the $\Lambda\Lambda$-$\Xi$N 
coupling effect in $^5_{\Lambda\Lambda}$He and 
$^5_{\Lambda\Lambda}$H, where energy of the $\Xi$ intermediate state is 
lowered by the formation of an $\alpha$ particle, which increases the 
coupling effect. \cite{10} The net change of the coupling
 effect in five-body hypernuclear systems consists of two effects,  
 $\Delta V_{\rm Pauli}$ and 
 $\Delta V_{\rm alpha}$, where the former is the Pauli suppression 
 effect, while the latter is an enhancement effect due to the formation 
 of an alpha particle in the $\Xi$ channel. These two effects are 
 written as 
 follows: 
\begin{equation}
\Delta V_{\rm Pauli} (^5_{\Lambda\Lambda}{\rm H}) = 
V_{\Lambda\Lambda,\Xi^0 {\rm n}}\frac {P_{\rm t}}{\Delta M_{1}} 
V_{\Xi^0 {\rm n},\Lambda\Lambda}
+\frac 12 (V_{\Lambda\Lambda,\Xi^- {\rm p}}\frac {P_{\rm t}}{\Delta M_{1}}
 V_{\Xi^- 
{\rm p},\Lambda\Lambda}),
\end{equation}
where
\begin{equation}
\Delta 
M_1 = \frac {M_{\Xi^0} + M_{\Xi^-}}2 + \frac {M_{\rm p} + M_{\rm n}} 2
- 2 M_{\Lambda}
 + 2 B^{\rm av}_{\Lambda} (^4_{\Lambda}{\rm H}) = 28.29~{\rm MeV}, 
 \end{equation}
and
\begin{equation}
\Delta V_{\rm alpha} (^5_{\Lambda\Lambda}{\rm H}) = \frac 12 
(V_{\Lambda\Lambda,\Xi^- {\rm p}}\frac {P_{\rm t}}{\Delta M_{1}}
 V_{\Xi^- 
{\rm p},\Lambda\Lambda}) - \frac 12 (V_{\Lambda\Lambda,\Xi^- {\rm p}}
\frac {P_{\alpha}}{\Delta M_{2}}
 V_{\Xi^- 
{\rm p},\Lambda\Lambda})
\end{equation}
with
\begin{equation}
\Delta 
M_2 = M_{\Xi^-} + M_{\alpha} - M_{\rm t} - 2 M_{\Lambda}
 + 2 B^{\rm av}_{\Lambda} (^4_{\Lambda}{\rm H}) =11.04~{\rm MeV}. 
 \end{equation}
 $P_{\rm t}$ and $P_{\alpha}$ are the projection operators on the 0$s$ 
 states of $^3$H and $^4$He, respectively, and
  $B_{\Lambda}¥^{\rm av}$ is the average 
 of the ground $0^+$ and the excited $1^+$ levels of $^4_{\Lambda}$He 
 or $^4_{\Lambda}$H as
 \begin{equation}
B^{\rm av}_{\Lambda} = \frac 
14B_{\Lambda}(0^+) + \frac 34
 B_{\Lambda}(1^+).
 \end{equation} 
 The 
 second term in $\Delta V_{\rm alpha}$ 
 corresponds to the coupling effect where the converted 
 proton is in the 0$s$ state and forms an alpha particle with the triton 
 core nucleus. 
 The first term is used to subtract an effect which is already included 
 in the single-channel $\Lambda\Lambda$ interaction. 
 Since the value of $\Delta M_2$ is smaller than that of 
$\Delta M_1$, $\Delta V_{\rm alpha}$ gives an attractive effect. 

Similarly, in the case of $^5_{\Lambda\Lambda}$He,    
\begin{equation}
\Delta V_{\rm Pauli} (^5_{\Lambda\Lambda}{\rm He}) = 
V_{\Lambda\Lambda,\Xi^- {\rm p}}
\frac {P_{\rm h}}{\Delta M_{1}} V_{\Xi^-{\rm p},\Lambda\Lambda}
+\frac 12 (V_{\Lambda\Lambda,\Xi^0 {\rm n}}\frac {P_{\rm h}}{\Delta M_{1}}
 V_{\Xi^0 
{\rm n},\Lambda\Lambda})
\end{equation}
with 
\begin{equation}
\Delta 
M_1 = \frac {M_{\Xi^0} + M_{\Xi^-}}2 + \frac {M_{\rm p} + M_{\rm n}} 2 
- 2 M_{\Lambda}
 + 2 B^{\rm av}_{\Lambda} (^4_{\Lambda}{\rm He}) = 28.83~{\rm MeV}, 
 \end{equation}
and
\begin{equation}
\Delta V_{\rm alpha} (^5_{\Lambda\Lambda}{\rm He}) = \frac 12 
(V_{\Lambda\Lambda,\Xi^0 {\rm n}}\frac {P_{\rm h}}{\Delta M_{1}}
 V_{\Xi^0 
{\rm n},\Lambda\Lambda}) - 
\frac 12 (V_{\Lambda\Lambda,\Xi^0 {\rm n}}\frac {P_{\alpha}}{\Delta M_{2}}
 V_{\Xi^0 
{\rm n},\Lambda\Lambda})
\end{equation}
with
\begin{equation}
\Delta 
M_2 = M_{\Xi^0} + M_{\alpha} - M_{\rm h} - 2 M_{\Lambda}
 + 2 B^{\rm av}_{\Lambda} (^4_{\Lambda}{\rm He}) = 5.69~{\rm MeV}, 
 \end{equation}
where $P_{\rm h}$ is the projection operator on the 0$s$ states 
of $^3{\rm He}$.
  In $^5_{\Lambda\Lambda}$He,
   $\Delta V_{\rm alpha}$ is more attractive 
   than that of  $^5_{\Lambda\Lambda}$H
   due to the smaller value of $\Delta 
  M_2$, which can be compared between eqs. (11) and (16). 
 
In our calculations, we use  
the fitted single-channel $\Lambda\Lambda$ interactions 
 and the averaged values of our $\Lambda$-h and 
$\Lambda$-t potentials, which are as follows:
\begin{equation}
V_{\Lambda-{\rm h/t}} = \frac 14 V_{\Lambda-{\rm h/t}}(0^+) 
+ \frac 34 V_{\Lambda-{\rm h/t}}(1^+).
\end{equation}
It is to be noted that 
$\Delta B_{\Lambda\Lambda}$ for 
the five-body system is 
defined by 
\begin{equation}
\Delta B_{\Lambda\Lambda} = 
B_{\Lambda\Lambda} - 
2B^{\rm av}_{\Lambda}
\end{equation}
with $B^{\rm av}_{\Lambda}$ of eq. (13) 
as was reasonably introduced by Filikhin and Gal. \cite{4} 
The results are displayed in table 4.

\begin{table}[htb]
 \begin{center}
 \caption{\label{tab:results1}
 Calculated -$\Delta B_{\Lambda\Lambda}$ for different
  $\Lambda\Lambda$ interactions in units of MeV.}
 \begin{tabular}{cccc}\\
\hline \hline  \\
 & $^6_{\Lambda\Lambda}{\rm He}$ &
   $^5_{\Lambda\Lambda}{\rm He}$ &   
  $^5_{\Lambda\Lambda}{\rm H}$ \\ \\
\hline \\ 
 $V_{\Lambda\Lambda}^{\rm D}$ & -1.48 & -1.13  & -1.07 \\ \\
 $ V_{\Lambda\Lambda}^{\rm D1}$ & -1.01 & -0.76 & -0.73 \\ \\
 $V_{\Lambda\Lambda}^{\rm D2}$ + $\Delta V_{\rm Pauli}^{\rm D}$ + 
 $\Delta V_{\rm alpha}^{\rm D}$ 
 & -1.01 & -0.86 & -0.77 \\ \\
 $V_{\Lambda\Lambda}^{\rm D2}$  & -1.10 & -0.83 & -0.79  \\ \\
 $\Delta V_{\rm Pauli}^{\rm D}$ &0.09 &0.05 & 0.05\\ \\
 $\Delta V_{\rm alpha}^{\rm D}$ &$-$ &-0.08 & -0.03\\ \\
\hline \\ 
$V_{\Lambda\Lambda}^{\rm e}$ &-0.64 &-0.47 &-0.45  \\ \\
 $ V_{\Lambda\Lambda}^{\rm e1}$ & -1.01 & -0.76 & -0.72 \\ \\
 $V_{\Lambda\Lambda}^{\rm e2}$ + $\Delta V_{\rm Pauli}^{\rm e}$ 
 +$\Delta V_{\rm alpha}^{\rm e}$ & -1.01 & -1.28 & -0.96 \\ \\
 $V_{\Lambda\Lambda}^{\rm e2}$  & -1.44 & -1.10 & -1.03 \\ \\
 $\Delta V_{\rm Pauli}^{\rm e}$ &0.43 & 0.23& 0.20\\ \\
 $\Delta V_{\rm alpha}^{\rm e}$ &$-$ &-0.41 & -0.13\\ \\
\hline \\ 
$V_{\Lambda\Lambda}^{\rm F}$ &-0.27 &-0.20 &-0.20  \\ \\
 $ V_{\Lambda\Lambda}^{\rm F1}$ & -1.01 & -0.74 & -0.70 \\ \\
 $V_{\Lambda\Lambda}^{\rm F2}$ + $\Delta V_{\rm Pauli}^{\rm F}$ 
 +$\Delta V_{\rm alpha}^{\rm F}$ & -1.01 & -1.92 & -1.20 \\ \\
 $V_{\Lambda\Lambda}^{\rm F2}$  & -1.89 & -1.45 & -1.36 \\ \\
 $\Delta V_{\rm Pauli}^{\rm F}$ &0.88 &0.47 & 0.43\\ \\
 $\Delta V_{\rm alpha}^{\rm F}$ &$-$ &-0.94 & -0.27\\ \\ 
\hline \hline 
 \end{tabular} 
 \end{center}
 \end{table}


The $\Lambda\Lambda$ interaction, 
$V_{\Lambda\Lambda}^{\rm e1}$, produces   
binding energy values, $\Delta B_{\Lambda\Lambda}$, of 0.72 MeV 
for $^5_{\Lambda\Lambda}$H and 0.76 MeV for 
$^5_{\Lambda\Lambda}$He as seen in table 4. 
The corresponding values produced by the 
free-space $\Lambda\Lambda$ interaction $V_{\Lambda\Lambda}^{\rm e2}$ 
are 1.03 MeV and 1.10 MeV, respectively. 
We then include the coupling 
effects $\Delta V_{\rm Pauli}^{\rm e}$ and $\Delta V_{\rm alpha}^{\rm e}$ 
in our calculations with $V_{\Lambda\Lambda}^{\rm e2}$. 
In $^5_{\Lambda\Lambda}$H, the contribution of the Pauli effect 
is 0.20 MeV and that of the $\alpha$ enhancement effect 
is -0.13 MeV for the NSC97e coupling strength. 
 Then, the net effect is a weak repulsion, which reduces 
the value of $\Delta B_{\Lambda\Lambda}$ in $^5_{\Lambda\Lambda}$H by 
0.07 MeV.

\vspace{-0.0cm}
\begin{figure}[htb]
\centerline{\epsfig{file=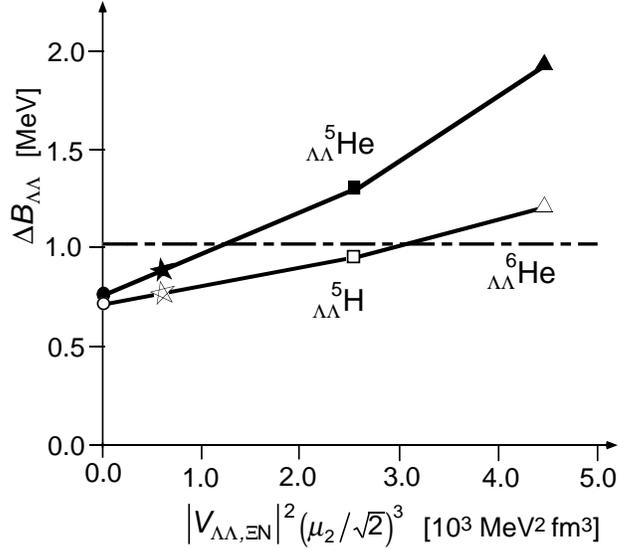,width=12.5cm}}
\vspace{-8.5cm}
\caption{$\Delta B_{\Lambda\Lambda}$ values against the coupling 
strength. The circles, stars, squares and triangles represent the cases of 
no, the NHC-D, the NSC97e 
and the NHC-F coupling 
strengths, respectively.}
\label{figXi}
\end{figure}
 
It is interesting to find that these effects 
give $\Delta B_{\Lambda\Lambda}$ = 1.28 MeV of
  $^5_{\Lambda\Lambda}$He, which is larger than 1.01 MeV
  of  $^6_{\Lambda\Lambda}$He, since the attractive 
  effect of $\Delta V_{\rm alpha}^{\rm e}$ 
  becomes larger than the suppression effect 
 of $\Delta V_{\rm Pauli}^{\rm e}$ in  $^5_{\Lambda\Lambda}$He. 
 The contributions of $\Delta V_{\rm Pauli}^{\rm e}$ and 
 $\Delta V_{\rm alpha}^{\rm e}$ 
 in this system are 0.23 MeV and -0.41 MeV.     
 Thus, the value of $\Delta B_{\Lambda\Lambda}$ for 
 $^5_{\Lambda\Lambda}$He 
 is increased by the coupling effects, while they 
 reduce the $\Delta B_{\Lambda\Lambda}$ value 
 of $^5_{\Lambda\Lambda}$H. 
 It should be noted that the value -0.41 MeV 
 of $\Delta V^{\rm e}_{\rm alpha}$ is a significant factor not to be 
 neglected in comparison with the empirical value 
 $\Delta B_{\Lambda\Lambda}$=1.01 MeV of $^6_{\Lambda\Lambda}$He.  
 Filikhin and Gal predicted the values of $\Delta B_{\Lambda\Lambda}$ 
 for $^5_{\Lambda\Lambda}$He and $^5_{\Lambda\Lambda}$H, which are 
 smaller than that of $^6_{\Lambda\Lambda}$He, and gave a comment that 
 ``the heavier is the core nucleus, the larger is the $\Delta 
 B_{\Lambda\Lambda}$.'' This  
 contradicts our results, where $\Delta B_{\Lambda\Lambda}$ of 
$^5_{\Lambda\Lambda}$He is larger than that of $^6_{\Lambda\Lambda}$He.

The values of $\Delta B_{\Lambda\Lambda}$ for the 
three double-$\Lambda$ hypernuclei are plotted against the coupling 
strength in fig. 2. With the value of $\Delta B_{\Lambda\Lambda}$
$(^6_{\Lambda\Lambda}$He) 
fixed to the experimental value of 1.01 MeV, it can be seen that the 
values for the five-body double-$\Lambda$ hypernuclei increase 
with the coupling strength. Thus, the coupling strength can be sensitively 
deduced from experimental observations of $^5_{\Lambda\Lambda}$He 
and $^5_{\Lambda\Lambda}$H.

\section{Conclusions}

\vspace{-0.15cm}

We have derived the  single-channel 
 $\Lambda\Lambda$ interactions 
based on Shinmura's $S$ = -2 interactions,
 which are phase-shift equivalents to the 
Nijmegen soft-core NSC97e, hard-core NHC-D and NHC-F models. 
The $\Lambda\Lambda$ 
interaction $V_{\Lambda\Lambda}^{\rm e1}$ is obtained by 
fitting the recent $^6_{\Lambda\Lambda}$He 
experimetal data, $\Delta B_{\Lambda\Lambda}$ = 1.01 MeV, \cite{3} without 
including the Pauli suppression effect. To obtain the free-space 
$\Lambda\Lambda$ interaction, however, 
we have to take into account the Pauli suppression 
effect in fitting the data, since it is  
appreciably large.
We have found that the Pauli 
suppression effect in $^6_{\Lambda\Lambda}$He is 0.43 MeV for the 
 NSC97e coupling strength. 
A $\Lambda$-$\Lambda$ phase shift of $12^{\circ}$ at the maximum 
 is produced by 
 the $\Lambda\Lambda$ interaction $V_{\Lambda\Lambda}^{\rm e1}$, 
 while the free-space  $\Lambda\Lambda$ interactions, 
 $V_{\Lambda\Lambda}^{\rm D2}$, 
 $V_{\Lambda\Lambda}^{\rm e2}$ and $V_{\Lambda\Lambda}^{\rm F2}$, 
 give $14^{\circ}$, 
 $17^{\circ}$ and $22^{\circ}$ maximum values, respectively. 
  Thus, the free-space $\Lambda\Lambda$ 
 interaction of the NSC97e and NHC-F cases 
 is considerably stronger than the interaction which is 
 obtained by fitting the Nagara event data while neglecting the Pauli 
 suppression effect.

 The coupling effects in the five-body systems consist of the Pauli 
 suppression, $\Delta V_{\rm Pauli}$, and an enhancement, 
 $\Delta V_{\rm alpha}$, which arises when a converted nucleon combines 
 with the core nuclues to form an $\alpha$ 
 particle. 
These two effects are largely 
 cancelled by each other in $^5_{\Lambda\Lambda}$H, and the resultant 
 effect is a repulsion with 0.07 (0.16) MeV for the NSC97e (NHC-F) coupling 
 strength. In $^5_{\Lambda\Lambda}$He, however, 
 the enhancement effect dominates, and the net 
 coupling effect is 
 not repulsion but 
  0.18 (0.47) MeV attraction for the NSC97e (NHC-F) coupling 
 strength. 
 The behaviour of $\Delta B_{\Lambda\Lambda}$ 
values against the coupling strength, shown in fig. 2, indicates the 
significance of the $\Lambda\Lambda$-$\Xi$N coupling effect,  
and observations of $^5_{\Lambda\Lambda}$He 
and $^5_{\Lambda\Lambda}$H would be critical for determining the 
 $\Lambda\Lambda$-$\Xi$N coupling strength.

 One possible way to produce $^5_{\Lambda\Lambda}$H through the
 ${\rm (K^-,K^+)}$ reaction has been discussed by Kumagai-Fuse and 
 Akaishi. \cite{15} They proposed that 
  $^5_{\Lambda\Lambda}$H is almost exclusively formed with a large 
  branching of about 90\%, once $^ 7_{\Xi}$H is populated by 
  the ${\rm (K^-,K^+)}$ reaction on a $^7{\rm Li}$ target. This process to 
  produce $^5_{\Lambda\Lambda}$H would be very promising if an intense 
  K$^-$ beam becomes available.
  
\section*{Acknowledgements}
The authors thank Dr. H. Nemura and Prof. M. Kohno for valuable 
discussions. One of the authors (Khin Swe Myint) likes to express her 
gratitude towards ``Japan Society for Promotion of Science'' (JSPS) for 
the fellowship grant during her stay in KEK, Japan where this work 
has been performed.  

\setcounter{equation}{0}
\def\theequation{A.\arabic{equation}}
\section*{Appendix A}

The total wave function of a three-body system  with 
masses $m_1$, $m_2$ and $m_3$ is expanded in terms  of Gaussian basis 
functions, which are spanned over the three rearrangement channels in 
the Jacobi coordinates ${\vec r}_c$ = ${\vec r}_i$ - ${\vec r}_j$ and 
 ${\vec R}_c$ = ${\vec r}_k$ 
 - $(m_i{\vec r}_i + m_j{\vec r}_j)/(m_i + m_j)$ as

\begin{equation}
 \Psi = \sum_{c=1}^3\sum_{i,j}^N A_{ij}^c e^{-(\frac {r_c}{b_i^c})^2}
e^{-(\frac {R_c}{b_j^c})^2},
\end{equation}
where 
\begin{equation}
b_{i+1} = cb_i.
\end{equation}
Typically, we use $b_1$ = 0.2 fm, $c$ = 1.4 and $N$ = 11 $\sim $ 15.
 The three-body Hamiltonian is 
 \begin{equation}
 H = -\frac {\hbar^2}{2\mu_c}\nabla^2_{r_c}
 -\frac {\hbar^2}{2M_c}\nabla^2_{R_c}
 + V_{12} + V_{23} +
 V_{31},
 \end{equation}
 where $\mu_c$ and $M_c$ 
 are the reduced masses corresponding to the Jacobi coordinates.
 $V_{ij}$'s are angular-momentum dependent potentials 
 with Gaussian-form radial parts,
 \begin{equation}
 \langle {\vec r}'\mid V \mid \vec r \rangle = 
 \sum_l V_l e^{-(\frac r{\mu})^2} \frac {\delta (r'-r)}{r^2} \sum_m
 Y_{lm}^*(\hat{\bf r}')  Y_{lm}(\hat{\bf r}).
 \end{equation} 
 The matrix elements of the potentials are reduced 
 to the following form:
 \begin{equation}
 4\pi (\frac{\pi}{C})^{\frac 32}\sum_{l}(2l+1)V_l
 {\int_0^{\infty} dr r^2 e^{-Ar^2} i^l j_{l}(-iBr^2)},
 \end{equation}
 where constants $A$, $B$ and $C$ are related 
 to the Gaussian-basis parameters of 
 the wave function and range parameters of the potentials.
The integral part of the above matrix element is evaluated to be
\begin{equation}
\frac 12 {\sqrt {\frac {\pi}2}} \frac {B^l}{\mid B\mid 
^{2l+1}} \frac {(A-\sqrt {A^2-B^2})^{l+\frac 12}} {\sqrt 
{A^2-B^2}}.
\end{equation}

We checked the accuracy of our three-body calculation 
on the $\Lambda$+$\Lambda$+$\alpha$ 
system with $V_{\Lambda\Lambda}$(97e) and $V_{\Lambda\alpha}$ from 
Filikhin and Gal. \cite{4} Our calculated value of $B_{\Lambda\Lambda}$, 
including all partial-wave contributions, 
is 6.90 MeV, which is in good agreement 
with 6.90 MeV of Nemura's variational method. \cite{16}
When only the $s$-state interactions 
are projected out, our calculation gives $B_{\Lambda\Lambda}$ = 6.70 
MeV, while that of Filikhin and Gal is 6.82 MeV. 
Thus, there is a discrepancy 
between our calculation and theirs; but, in the case of 
$V_{\Lambda\Lambda}$ = 0, our calculated value of 
$B_{\Lambda\Lambda}$ is 6.27 MeV, which coincides with their value. 
From our calculation, 
it is found that the 
higher partial wave contribution to the binding energy of 
$^6_{\Lambda\Lambda}$He is 0.20 MeV when all the interactions are 
taken to be of $s$-state. However, when we apply less attractive 
$p$-state $\Lambda$-$\alpha$ potential to the odd states, the 
contribution reduces to 0.14 MeV.

Non-local potentials $\Delta V_{\rm Pauli}$ and $\Delta V_{\rm alpha}$ 
are included in our calculations by
\begin{equation}
\langle \vec r'\vec R'\mid \frac P{\Delta M} \mid \vec r \vec R \rangle =
(\frac a{\pi})^3 e^{-\frac 14 a r'^2} e^{-a R'^2} \frac 1{\Delta M} 
e^{-\frac 14 a r^2} e^{-a R^2},
\end{equation}
where the harmonic-oscillator strength is taken to be 
$a$ = 0.521 fm$^{-2}$ for $^4$He   
and $a$ = 0.386 fm$^{-2} $ for $^3$He and $^3$H.

\end{document}